\documentclass{aa}
\usepackage{txfonts}
\usepackage{graphicx}
\usepackage{mathrsfs}

\usepackage[usenames]{color}

\newcommand{\cch}{$\mathrm{C_{2}H}$}
\newcommand{\nnh}{$\mathrm{N_{2}H^{+}}$}

\newcommand{\be}{\begin{equation}}
\newcommand{\ee}{\end{equation}}

\begin{document}

\title{Hydrogen cyanide and isocyanide in prestellar cores\thanks{Based on observations carried out with the IRAM 30m Telescope. IRAM is supported by INSU/CNRS (France), MPG (Germany) and IGN (Spain).}}

\author{M. Padovani\inst{1,2}, C.~M. Walmsley\inst{2,3}, M. Tafalla\inst{4}, 
P. Hily-Blant\inst{5} and G. Pineau des For\^ets\inst{6,7}}
\authorrunning{M. Padovani et al.}

\institute{
Institut de Ci\`encies de l'Espai (CSIC--IEEC), Campus UAB, Facultat de Ci\`encies, Torre C5--parell 2$^{a}$, E--08193 Bellaterra, Spain\\
\email{padovani@ieec.uab.es}
\and 
INAF--Osservatorio Astrofisico di Arcetri, Largo E. Fermi 5, I--50125 
Firenze, Italy
\and
Dublin Institute of Advanced Studies, 31 Fitzwilliam Place, Dublin, Ireland
\and
Observatorio Astron\'omico Nacional (IGN), Alfonso XII 3, E--28014 Madrid, 
Spain
\and
LAOG (UMR 5571) Observatoire de Grenoble, BP 53 F--38041 Grenoble Cedex 9, France
\and
IAS (UMR 8617), Universit\'e de Paris-Sud, F--91405 Orsay, France
\and
LERMA (UMR 8112), Observatoire de Paris, 61 Avenue de l'Observatoire, F--75014, Paris, France
}


\abstract {} 
{We studied the abundance of HCN, H$^{13}$CN, and HN$^{13}$C in a
sample of prestellar cores, in order to search for species associated with high
density gas.}
{We used the IRAM 30m radiotelescope to observe along the major and
the minor axes of \object{L1498}, \object{L1521E}, and \object{TMC 2}, three cores chosen on the basis of
their CO depletion properties. We mapped the $J=1\rightarrow0$ transition of
HCN, H$^{13}$CN, and HN$^{13}$C towards the source sample plus the
$J=1\rightarrow0$ transition of N$_2$H$^+$ and the $J=2\rightarrow1$ transition of
C$^{18}$O in \object{TMC 2}. We used two different radiative transfer codes,
making use of recent
collisional rate calculations, in order
to determine more accurately the excitation temperature, leading to a
more exact evaluation of the column densities and abundances.}
{We find that the optical depths of both H$^{13}$CN(1$-$0) and HN$^{13}$C(1$-$0) are
non-negligible, allowing us to estimate excitation temperatures for
these transitions in many positions in the three sources.  
The observed excitation temperatures are consistent with recent 
computations of the collisional rates for these species and they
correlate with hydrogen column density inferred from dust emission.
We conclude that HCN and HNC are relatively abundant in the high
density zone, $n({\rm H}_{2})\sim10^5$ cm$^{-3}$, where CO is depleted. 
The relative abundance [HNC]/[HCN]
differs from unity by at most 30\% consistent with chemical
expectations. The three hyperfine satellites
of HCN(1$-$0) are optically thick in the regions mapped, but the profiles
become increasingly skewed to the blue (\object{L1498} and \object{TMC 2}) or red (\object{L1521E})
with increasing optical depth suggesting absorption by foreground 
layers.}
{}


\maketitle

\section{Introduction}

 Tracing the dense central regions of prestellar cores is essential
 for an understanding of their kinematics and density distribution.
 Many of the well studied nearby cores are thought to be static with
 a "Bonnor-Ebert" radial density distribution (see Bergin \&
Tafalla \cite{bt07} and references therein) comprising a roughly $1/r^2$ 
 fall off surrounding a central region of nearly constant
density.  However, our understanding of the central regions 
 with H$_{2}$ number densities of $10^5$ cm$^{-3}$ or more is limited
  both by our uncertainties about dust emissivity and molecular
abundances.  Dust emission, because it is optically thin, is a
useful tracer of mass distribution, but both temperature gradients
and gradients in the dust characteristics (refractive index and
size distribution) mean that dust emission maps should be interpreted
with caution. Molecular lines offer the advantage that they permit
an understanding of the kinematics, but depletion and other chemical
effects render them often untrustworthy.  Depletion is more rapid
at high density and low temperature and these are just the effects
that become important in the central regions of cores.  

 One interesting result from studies to date (e.g. Tafalla et al. \cite{tm06}) is 
that the only molecular species to have a spatial distribution 
similar to that of the optically thin dust emission are 
N$_{2}$H$^{+}$ and NH$_{3}$, consisting solely of nitrogen and hydrogen.  This is
in general attributed to the claim that their abundance is  closely
linked to that of molecular nitrogen (one of the main gas phase
repositories of nitrogen) which is highly volatile and hence one
of the last species to condense out (Bisschop et al. \cite{bf06}). 
However N$_{2}$ is only marginally more volatile than CO which is
found observationally 
to condense out at densities of a few 10$^{4}$ cm$^{-3}$
(Tafalla et al. \cite{tm02}, \cite{tm04}) and it has thus been puzzling to have a
scenario where C-containing species were completely frozen out
but N$_{2}$ not. This has led to a number of models aimed at 
retaining gas phase nitrogen for at least a short time
after CO has disappeared, e.g. Flower et al. \cite{fp06},
Akyilmaz et al. \cite{af07}, and Hily-Blant et al. \cite{hw10} (hereafter paper I). 
There have also
been attempts to obtain convincing evidence for 
the existence in the gas phase of species 
containing both C and N at densities above a few 10$^5$ to 10$^6$ cm$^{-3}$
at which CO depletes (e.g. Hily-Blant et al. \cite{hw08} and paper I).
A partial success in this regard was obtained by Hily-Blant et
al. \cite{hw08} who found that the intensity of $^{13}$CN(1$-$0) behaved in
similar fashion to the dust emission in two of the densest cores:
L183 and L1544.  The implication of this is that at least
some form of carbon, probably in the form of CO, remains in the
gas phase in these objects at densities above $10^5$ cm$^{-3}$.
Thus, a tentative conclusion is that
the CO abundance at densities above 10$^5$ cm$^{-3}$
is much lower than the canonical value
of $10^{-4}$ relative to H$_2$ (Pontoppidan \cite{p06}), 
but nevertheless sufficiently high to
supply carbon for minor species.

Putting this interpretation on a more solid foundation requires
observation of possible tracers of the depleted region in cores
carried out in a manner which will permit distinguishing species
associated with high density gas from those present in the surrounding
lower density envelope.  This is complicated though perhaps facilitated
by the gradients in molecular abundance due to depletion. While
molecular abundances in general are expected to drop at high 
densities when depletion takes over, this is not necessarily true
for all species at least within a limited density range (see the
results for deuterated species discussed by Flower et al. \cite{fp06}). 
However, proving that one is observing emission from high density
gas involves either showing that one can detect transitions which 
require high densities to excite or showing that the emission
comes from a compact region coincident with the dust emission
peak. The former is rendered more difficult by the temperature
gradients believed to exist in some cores (see e.g. Crapsi
et al. \cite{cc05}). The latter requires
highly sensitive high angular resolution observations.
A combination of the two is likely to be the best strategy.

 One approach to these problems which has not been fully exploited 
 to date is to make use of the hyperfine splitting present in
 essentially all low lying transitions of N-containing species.
  It is clear in the first place that the relative populations
 of hyperfine split levels of species such as HCN and
CN are out of LTE in many situations
 (see Walmsley et al. \cite{wc82}, Monteiro \& Stutzki \cite{ms86}, and paper I). 
 Interpreting such anomalies requires
accurate collisional rates between individual hyperfine split 
levels, but such rates  can be determined from the rates between
rotational levels (e.g. Monteiro and Stutzki \cite{ms86}) and clearly
in principle, they allow limits to be placed on the density of
the emitting region.  It is also the case that species of
relatively low abundance like H$^{13}$CN are found to have 
rather minor deviations from LTE between hyperfine levels of
the same rotational level (see paper I) and in this
case, one can  determine the line excitation temperature
and optical depth based on the relative hyperfine satellite 
intensities.  While this is clearly doubtful procedure in that
the non-LTE anomalies can  cause errors in the inferred optical
depth, it is nevertheless (as we shall discuss) an approach which can
give a zero order estimate of relative rotational level populations
 and hence of the local density.

In the case of more abundant species such as the main 
isotopologues of HCN or CN, one finds  that "self absorption" or 
absorption in foreground relatively low density material  can
obliterate any signal from the high density central regions
of a core.  However, from these tracers, one can in principle 
glean information on the kinematics of foreground layers of
the core.  Thus, one can hope to find evidence for infall. 
We indeed find in this study that profiles change in gradual
fashion as a function of transition line strength and this does
indeed yield evidence for either infall or expansion of foreground
layers. 

 Here we extend the results of paper I to three other
cores which have been chosen on the basis of their CO depletion 
properties (see e.g. Bacmann et al. \cite{bl02}, Brady Ford \& Shirley \cite{bs11}).  
We in particular chose objects with large CO depletion
holes indicative of a relatively large age.  This might occur
for example in sources where magnetic field is capable of slowing
down or preventing collapse. We also used as a guide the angular
size measured in the N$_2$H$^+$(1$-$0) transition by Caselli et al. (\cite{cm02})
which is thought to correspond roughly to the area of CO depletion.
Thus, we chose the cores \object{L1498} and \object{TMC 2} which appear to have large
CO depletion holes (0.08 parsec in the case of \object{TMC 2}). As a comparison source,
we included in our study \object{L1521E}, which shows little or no depletion
in C-bearing species (Tafalla \& Santiago \cite{ts04}). 

In this paper then, we present IRAM 30m observations of 
the $J=1\rightarrow0$ transition of HCN, H$^{13}$CN and HN$^{13}$C along the major
and the minor axes of the three selected sources, \object{L1498}, \object{L1521E} and \object{TMC 2}, as well as
the $J=1\rightarrow0$ transition of \nnh\ and the
$J=2\rightarrow1$ transition of C$^{18}$O towards \object{TMC 2}. 
In Sect. \ref{HCN:observations}, we
discuss the observational and data reduction procedures, summarising the results in
Sect. \ref{HCN:obsresults}. 
In Sect.~\ref{HCN:depletion} we compare the distribution of the line emission with
respect to the dust emission, investigating about the presence of depletion. 
In Sect.~\ref{HCN:lineprofiles} we describe 
the evidence of the asymmetry of the HCN(1$-$0)
line profiles. In Sect.~\ref{HCN:Texestimate} we explain the methods used for
determining the excitation temperature, while in Sect.~\ref{HCN:Nx}
we give the
estimated column densities and abundances 
for the observed lines.
In Sect.~\ref{HCN:conclusions} we give our conclusions.
Comment on non-LTE hyperfine populations are provided in Appendix~\ref{nonltepop},
while in Appendix~\ref{app2} we summarise 
the spectroscopic data and observational parameters
of the observed molecules.

\section{Observations}\label{HCN:observations}
\object{L1498}, \object{L1521E}, and \object{TMC 2} were observed in July 2008, using frequency switching
and raster mode, with a spacing of 
20$^{\prime\prime}$,
25$^{\prime\prime}$ and 50$^{\prime\prime}$, respectively, along the major and
the minor axes. The axes were identified from the continuum
emission maps showed in Tafalla et al. (\cite{tm02}) for \object{L1498}, Tafalla et al. (\cite{tm04})
for \object{L1521E}, and Crapsi et al. (\cite{cc05}) for \object{TMC 2}. Observing parameters are
given in Table~\ref{tab:obspam}.


Observations of the 89 GHz HCN(1$-$0) and the 86 GHz H$^{13}$CN(1$-$0) multiplets (see 
Table \ref{tab:HCN10}) plus the 87 GHz HN$^{13}$C(1$-$0) multiplet
(see Table \ref{tab:HN13C10}) were carried out using the VESPA autocorrelator
with 20 kHz channel spacing (corresponding to about
0.069 km s$^{-1}$) with 40 MHz bandwidth for HCN(1$-$0) and 20 MHz bandwidth for
H$^{13}$CN(1$-$0) and HN$^{13}$C(1$-$0).
The final rms, in $T_{\rm mb}$ unit is $\sigma_{\rm T}\sim50$ mK, $\sim20$ mK,
and $\sim30$ mK for HCN(1$-$0), H$^{13}$CN(1$-$0), and HN$^{13}$C(1$-$0),
respectively.

Observations of the 93 GHz N$_{2}$H$^{+}$(1$-$0) multiplet and of the 219 GHz
C$^{18}$O(2$-$1) line were carried out in \object{TMC 2} using
10 kHz channel spacing with 40 MHz bandwidth for \nnh(1$-$0) and 20 kHz
channel spacing with 40 MHz bandwidth for C$^{18}$O(2$-$1).
For \nnh\, the final rms, in $T_{\rm mb}$ units, in channels of width $\delta v = 0.031$
km s$^{-1}$, is $\sigma_{\rm T}\sim100$ mK while for 
C$^{18}$O(2$-$1),
the final rms (channels of width $\delta v=0.027$ km s$^{-1}$) is
$\sigma_{\rm T}\sim 250$ mK. 

Data reduction and analysis were completed using the CLASS software 
of the GILDAS\footnote{\tt http://www.iram.fr/IRAMFR/GILDAS} facility,
developed at the IRAM and the Observatoire de Grenoble.
In what follows, all temperatures are on the main-beam scale, 
$T_{\rm mb}=F_{\rm eff}T_{\rm A}^{*}/B_{\rm eff}$, where $T_{\rm A}^{*}$ is the antenna
temperature corrected for atmospheric absorption, while B$_{\rm eff}$ and F$_{\rm eff}$
are the beam and the forward efficiencies, respectively
(see Table \ref{tab:obspam} for the numerical values of efficiencies).

\section{Observational results}\label{HCN:obsresults}
\begin{figure*}[!ht]
\begin{center}
\includegraphics[angle=0,scale=.75]{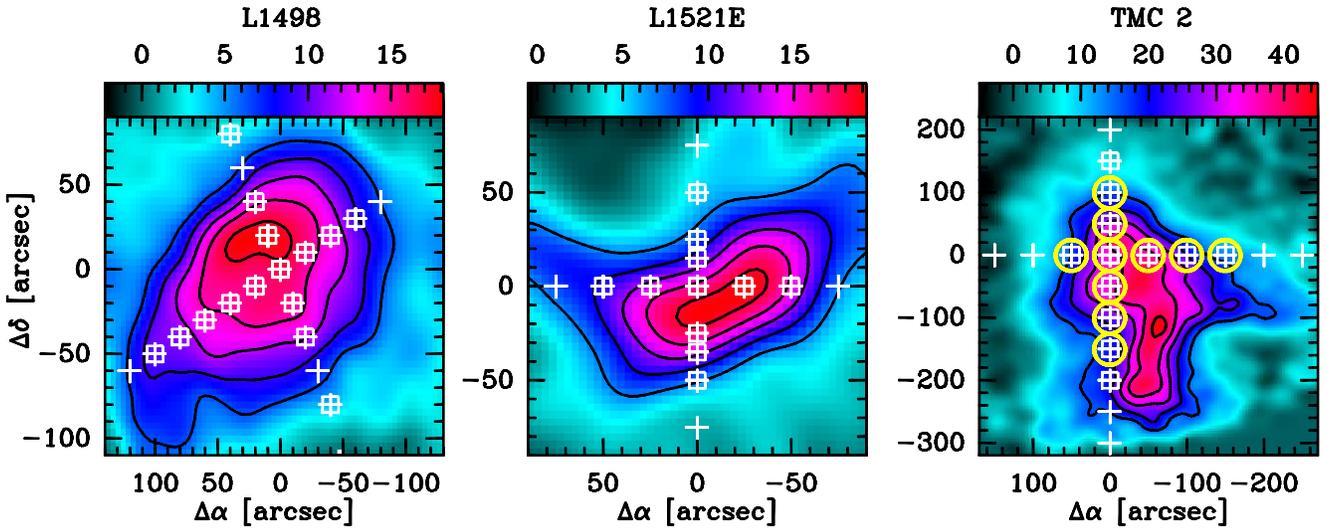}
\caption{Observed positions in \object{L1498} (left panel), \object{L1521E} (middle panel), and
\object{TMC 2} (right panel), superposed on the dust emission map 
smoothed to 28$^{\prime\prime}$, from Tafalla et al. (\cite{tm02}), Tafalla et al. (\cite{tm04}), and
Crapsi et al. (\cite{cc05}), respectively. Mapped positions in
HCN(1$-$0) ({\em white crosses}), 
H$^{13}$CN(1$-$0) and HN$^{13}$C(1$-$0) ({\em white squares}), and 
N$_{2}$H$^{+}$(1$-$0) ({\em yellow circles}). Contours represent 35, 50, 65, 80 and 95 
per cent of the peak value of the dust emission which is 18.0, 21.0, and 45.0 
mJy/(11$^{\prime\prime}$ beam) for \object{L1498}, \object{L1521E}, and \object{TMC 2}, respectively.
The (0,0) position corresponds to $\alpha(2000)=04^{\rm h}10^{\rm m}51.5^{\rm s}$, 
$\delta(2000)=25^\circ09^{\prime}58^{\prime\prime}$ for \object{L1498}, to 
$\alpha(2000)=04^{\rm h}29^{\rm m}15.7^{\rm s}$, 
$\delta(2000)=26^\circ14^{\prime}05^{\prime\prime}$ for \object{L1521E}, and to
$\alpha(2000)=04^{\rm h}32^{\rm m}48.7^{\rm s}$, 
$\delta(2000)=24^\circ25^{\prime}12^{\prime\prime}$ for \object{TMC 2}.}
\label{17134fg1}
\end{center}
\end{figure*}

Figure \ref{17134fg1} shows the continuum emission of the
three cores together with the positions mapped in lines described in Sect.
\ref{HCN:observations}.
The selected sources have already been widely observed in the past
as possible tracers of conditions in the high density
core nucleus.
We succeeded in observing HN$^{13}$C(1$-$0) and
the three hyperfine components of HCN(1$-$0) and H$^{13}$CN(1$-$0).

\begin{figure}[]
\begin{center}
\resizebox{\hsize}{!}{\includegraphics[angle=0]{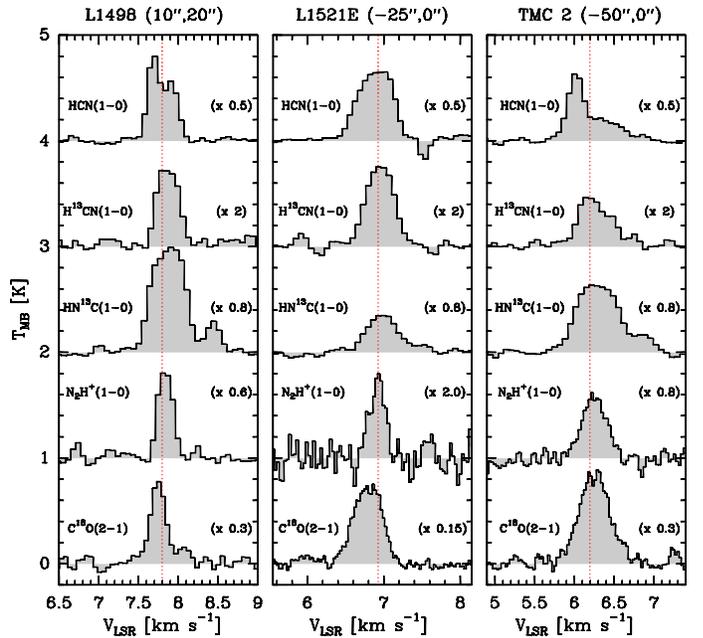}}
\caption{Emission line for the different tracers of the source sample towards the 
dust emission peak. The 
88633 MHz component of
HCN(1$-$0) is shown together with the 86340 MHz
component of H$^{13}$CN(1$-$0), HN$^{13}$C(1$-$0), the 93176 MHz
isolated component of
\nnh(1$-$0), and C$^{18}$O(2$-$1). Spectra have been multiplied by
a scaling factor to allow the simultaneous comparison. 
Red dotted lines show the
systemic LSR velocity of the sources evaluated from the well-determined frequencies
of \nnh(1$-$0), see Pagani et al. (\cite{pd09}).}
\label{17134fg2}
\end{center}
\end{figure}

Figure \ref{17134fg2} shows a comparison of the line profiles for the different
tracers. 
We 
plotted the weakest \mbox{($F=0\rightarrow1$)} component of
HCN(1$-$0) at 88633.936 MHz,
the strongest ($F=2\rightarrow1$) component of H$^{13}$CN(1$-$0) at
86340.184 MHz, HN$^{13}$C(1$-$0) at 87090.675 MHz, the isolated component 
($F_{1},F=0,1\rightarrow1,2$) of
\nnh(1$-$0) at 93176.2650 MHz, and the C$^{18}$O(2$-$1) transition at 219560.319 MHz.
We used the components enumerated here also for the comparison between line and
dust emission described below, supposing them to be the most
optically thin.
The offsets for each source are relative to the dust peak emission 
(see caption of Fig.~\ref{17134fg1}).

There are clear trends in the line widths which we derive for our
three sources with values of order 0.2 km s$^{-1}$ for \object{L1498}, 0.3 km s$^{-1}$
for \object{L1521E}, and 0.45 km s$^{-1}$ for \object{TMC 2}. 
Besides, we noticed that 
there is a reasonable agreement between
the systemic velocity, $V_{\rm LSR}$, of \nnh\ and H$^{13}$CN
to within 0.08 km s$^{-1}$, while
line widths of H$^{13}$CN
seem larger by 0.05 km s$^{-1}$ with respect to \nnh, 
but this may be due to $^{13}$C hyperfine splitting
(Schmid-Burgk et al. \cite{sm04}).
   
There is a clear difference between the HCN and H$^{13}$CN profiles
in the three sources: while in \object{L1521E}, the HCN profile is broader than its
isotopologue, in \object{L1498} and \object{TMC 2}, HCN and H$^{13}$CN have very different profiles,
presumably because of high optical depth in HCN. 
  
\begin{figure}[]
\begin{center}
\resizebox{\hsize}{!}{\includegraphics[angle=0]{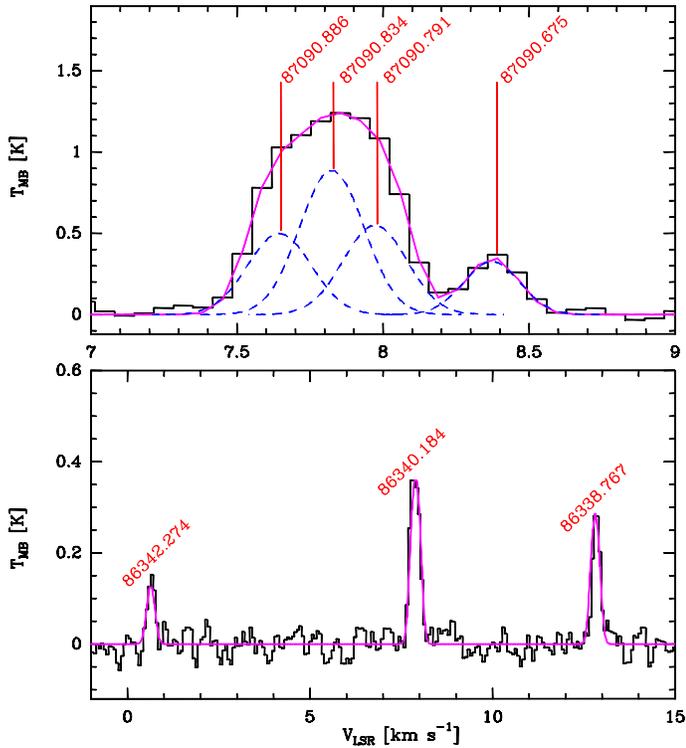}}
\caption{HN$^{13}$C(1$-$0) (upper panel) and
H$^{13}$CN(1$-$0) (lower panel)
emission towards \object{L1498} at the offset $(10,20)$;
{\em black solid histogram}, hyperfine components used for the fit,
{\em blue dashed lines}, centred at the frequencies listed in Table 
\ref{tab:HN13C10}, and line fit, {\em magenta solid 
line}.}
\label{17134fg3}
\end{center}
\end{figure}

\begin{figure}[]
\begin{center}
\resizebox{\hsize}{!}{\includegraphics[angle=0]{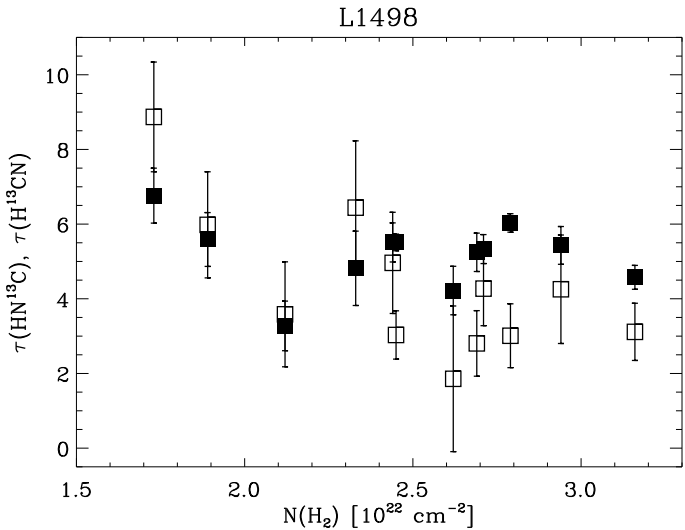}}
\caption{Total optical depth of HN$^{13}$C(1$-$0), {\em solid squares}, and 
H$^{13}$CN(1$-$0), {\em empty squares},
for all the observed positions in \object{L1498} as a function of
the molecular hydrogen column density.}
\label{17134fg4}
\end{center}
\end{figure}

All these molecules show hyperfine structure, although in the case of
HN$^{13}$C it is
not possible to avoid the blending of the components,
and the hyperfine fitting has been conducted using the HFS method in CLASS. 
To fit this line, we followed van der Tak et al. (\cite{vdtm09}) who found that 
it consists of eleven hyperfine components which can however be reduced to four
``effective'' components.
As an instance, in the upper panel of Fig.~\ref{17134fg3}, 
we show the fit of the HN$^{13}$C 
line at the offset $(10,20)$ towards \object{L1498} together with the four distinguishable
hyperfine components. 
The fit gives a value for the total optical depth equal to 
$\tau=4.58\pm0.32$ and all the observed points show similar values,
as shown in Figure~\ref{17134fg4}, with an 
average of $\langle\tau\rangle=5.16\pm0.86$.
Also 
for H$^{13}$CN we found a good simultaneous fit of the three hyperfine components
(see lower panel of Fig. \ref{17134fg3}) 
and, as HN$^{13}$C, this line is somewhat optically thick with a total optical depth of
$\tau=3.11\pm0.77$ at the offset $(10,20)$ and a mean optical depth of $\langle\tau\rangle=4.34\pm1.15$. 
Hence one can expect the optical depth in
HCN hyperfine components to be of order 30--100 assuming the canonical isotopic
ratio for $[^{12}$C]/$[^{13}$C] of 68 (Milam et al.~\cite{ms05}).

\section{Probing the presence of depletion}
\label{HCN:depletion}
\begin{figure}[!h]
\begin{center}
\resizebox{\hsize}{!}{\includegraphics[angle=0]{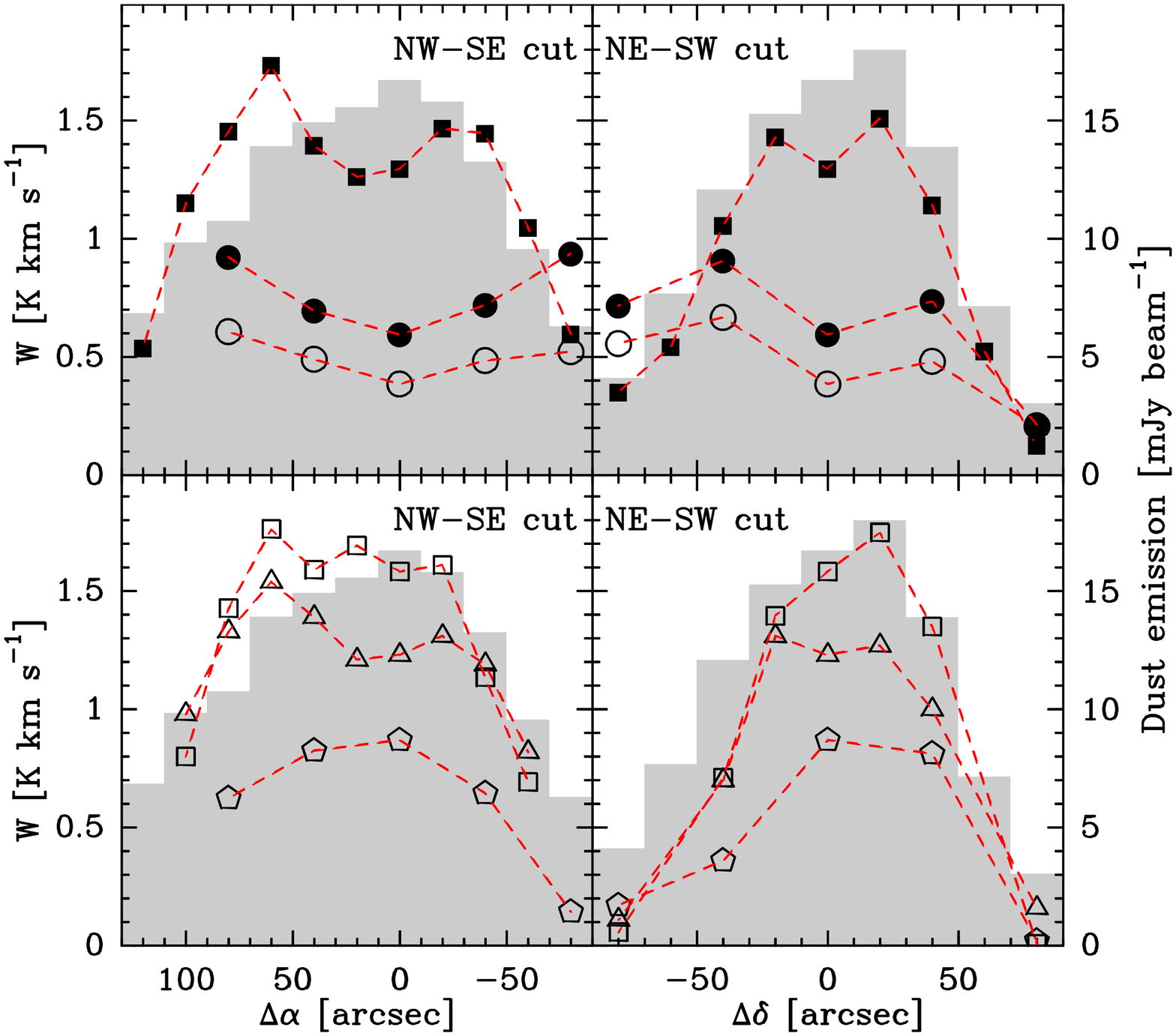}}
\caption{Comparison between the dust emission ({\em gray histograms}) and the integrated 
intensity of the observed species along the two cuts in \object{L1498}. Upper panels: 
{\em solid squares}, HCN(1$-$0) [$\times 3$]; 
{\em empty circles}, C$^{18}$O(1$-$0) [$\times0.8$]; 
{\em solid circles}, C$^{18}$O(2$-$1) [$\times1.4$]. 
Lower panels: {\em empty triangles}, H$^{13}$CN(1$-$0) [$\times10$]; 
{\em empty squares}, HN$^{13}$C(1$-$0) [$\times2.5$]; 
{\em empty pentagons}, N$_{2}$H$^{+}$(1$-$0) [$\times 3$].
The typical error on the integrated intensity is about 20 mK km s$^{-1}$.}
\label{17134fg6}
\end{center}
\end{figure}

\begin{figure}[!h]
\begin{center}
\resizebox{\hsize}{!}{\includegraphics[angle=0]{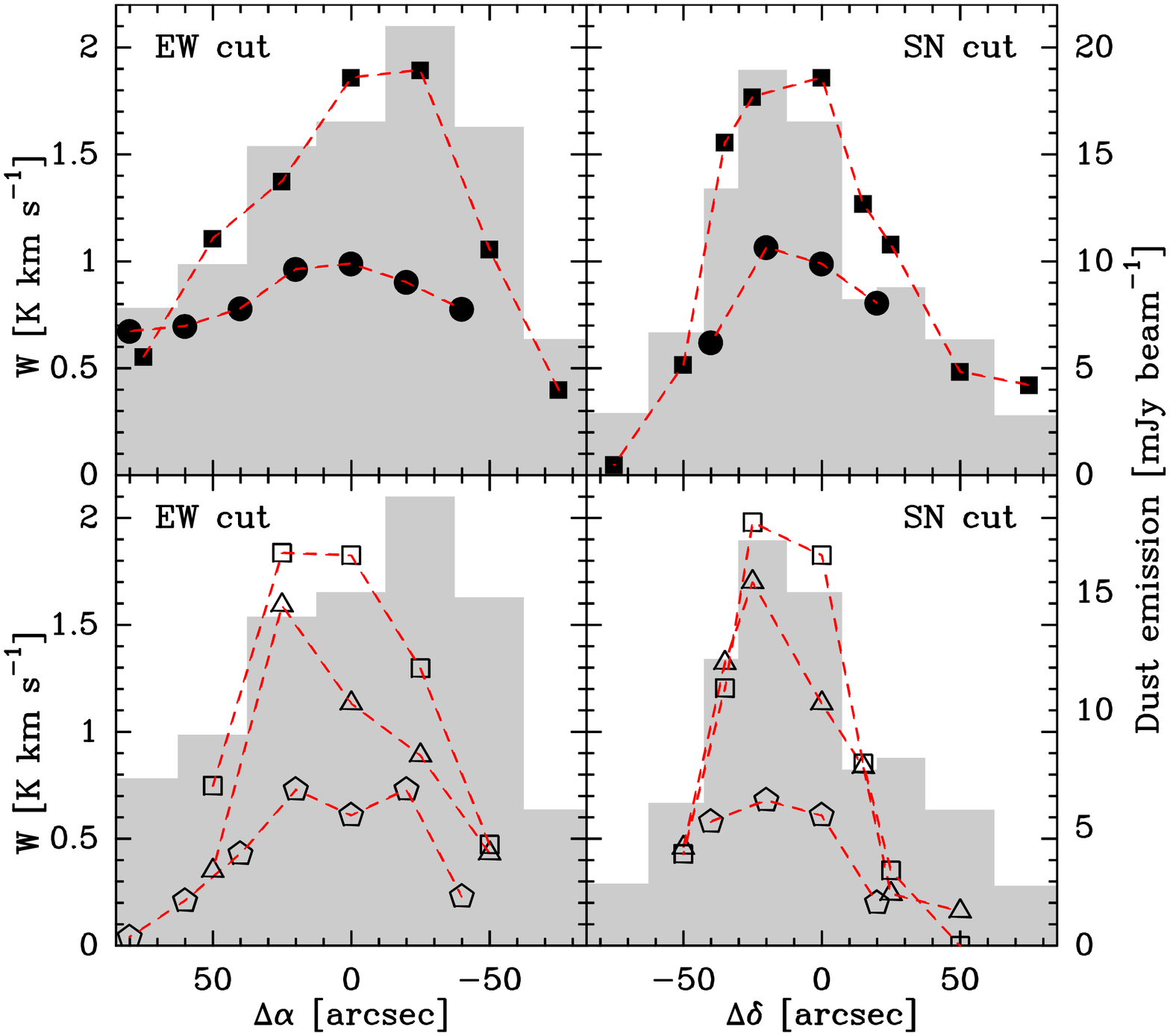}}
\caption{Comparison between the dust emission ({\em gray histograms}) and the integrated 
intensity of the observed species along the two cuts in \object{L1521E}. Upper panels: 
{\em solid squares}, HCN(1$-$0) [$\times 2.5$]; 
{\em solid circles}, C$^{18}$O(2$-$1) [$\times0.4$]. 
Lower panels: {\em empty triangles}, H$^{13}$CN(1$-$0) [$\times27$]; 
{\em empty squares} HN$^{13}$C(1$-$0) [$\times5.5$]; 
{\em empty pentagons}, N$_{2}$H$^{+}$(1$-$0) [$\times 10$].
The typical error on the integrated intensity is about 20 mK km s$^{-1}$.}
\label{17134fg7}
\end{center}
\end{figure}

\begin{figure}[!h]
\begin{center}
\resizebox{\hsize}{!}{\includegraphics[angle=0]{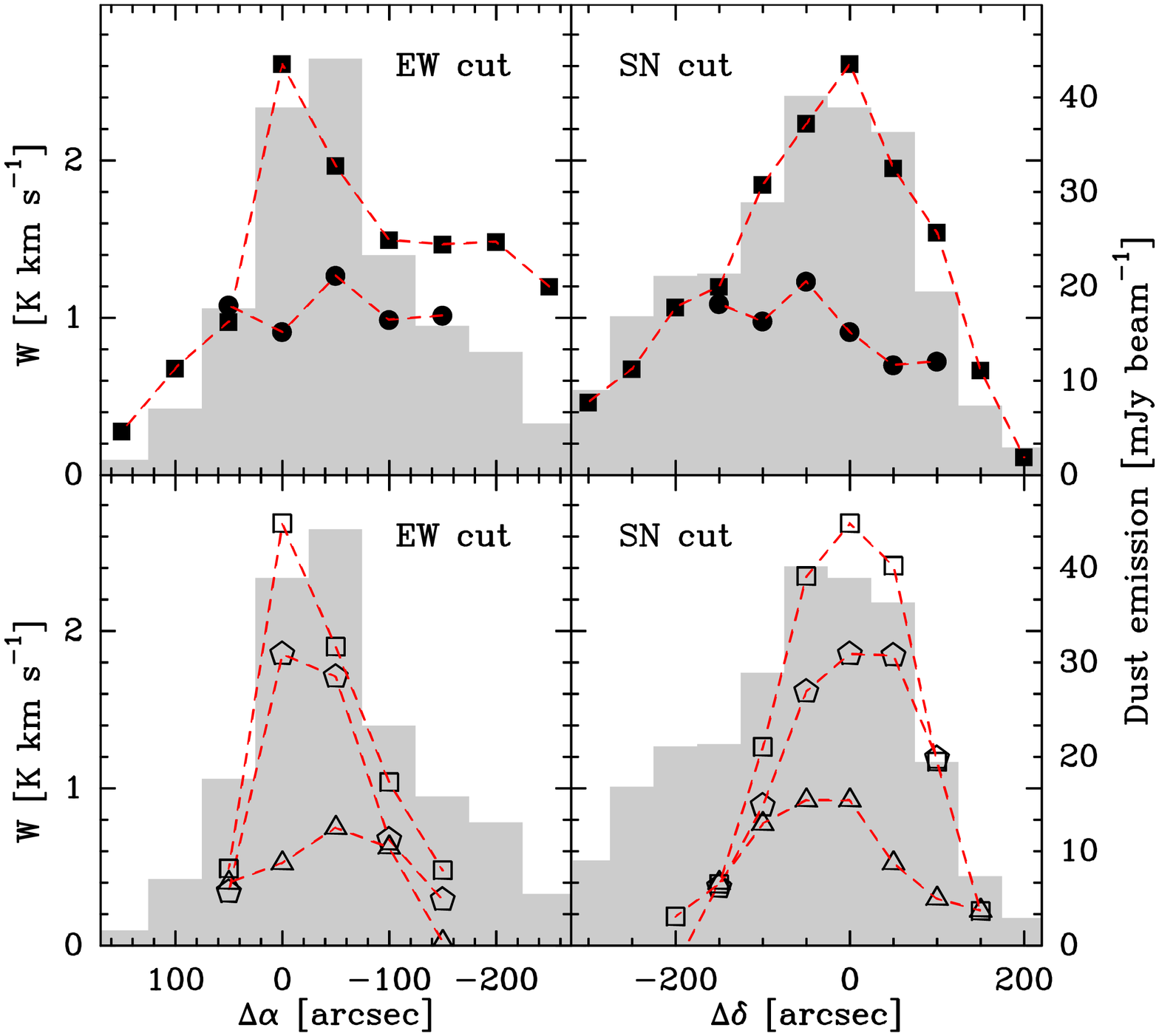}}
\caption{Comparison between the dust emission ({\em gray histograms}) and the 
integrated intensity of the observed species 
along the two cuts in \object{TMC 2}. Upper panels: 
{\em solid squares}, HCN(1$-$0) [$\times 4$];
{\em solid circles}, C$^{18}$O(2$-$1). 
Lower panels: {\em empty triangles}, H$^{13}$CN(1$-$0) [$\times25$]; 
{\em empty squares}, HN$^{13}$C(1$-$0) [$\times3.5$]; 
{\em empty pentagons}, N$_{2}$H$^{+}$(1$-$0) [$\times4.5$].
The typical error on the integrated intensity is about 20 mK km s$^{-1}$.}
\label{17134fg8}
\end{center}
\end{figure}

In this section, we study the dependence of the integrated 
intensity, $W$, of the HCN, H$^{13}$CN and HN$^{13}$C transitions
on offset from the dust emission peaks. In the case of HCN,
we consider the weakest $F=0\rightarrow1$ 
line and in the case of H$^{13}$CN, the
strongest $F=2\rightarrow1$ satellite. 
The observational results are shown in
Figures~\ref{17134fg6}, 
\ref{17134fg7}, 
and \ref{17134fg8} 
where we compare the observed intensity of these
lines with the dust emission which is taken to be representative of
the H$_{2}$ column density. It is useful first to comment briefly on 
what one can expect qualitatively to learn from such studies.

One notes in the first place that optically thick transitions,
due to scattering and line saturation, are naturally likely to
have a broader spatial distribution than thin (or very moderate
optical depth) transitions. Thus, the HCN($F=0\rightarrow1$)
transition
can be expected to be roughly an order of magnitude more optically
thick than H$^{13}$CN($F=2\rightarrow1$) 
assuming the local interstellar
[$^{12}$C]/[$^{13}$C] ratio 
(Milam et al. \cite{ms05}) and no fractionation. 
One
thus expects a broader spatial distribution of HCN than H$^{13}$CN and indeed this
is confirmed by the results shown in all the three sources.  For
example, the half power size of HCN in \object{L1498} along the NW-SE cut
is 170$^{\prime\prime}$ as
compared with 120$^{\prime\prime}$ for H$^{13}$CN
and HN$^{13}$C, 
and 180$^{\prime\prime}$ for the dust emission. One concludes
that the $^{13}$C substituted isotopologue is a much better tracer of HCN
column density than the more abundant form even when using the weakest
hyperfine component.

However, one also notes from these figures that the
C$^{18}$O distribution is essentially flat as has been seen in a variety
of studies (Tafalla et al. \cite{tm02}, \cite{tm04}). 
This has been attributed to depletion of CO at densities
$n$(H$_{2}$) above a critical value of order a few times 10$^{4}$ cm$^{-3}$ 
and our observations are
entirely consistent with this.   
It is also true however (see also Tafalla et al. \cite{tm06} and paper I)
that species like H$^{13}$CN and HN$^{13}$C,
while they clearly have different spatial distributions from the
dust emission, have half power sizes which are roughly similar
(see results for \object{L1498} above).
Such effects clearly do not apply to \nnh\ which does
not contain carbon and which in general 
has been found to have a spatial distribution similar to that 
of the dust (Tafalla et al. \cite{tm06}). 
However, one sees in \mbox{Fig.~\ref{17134fg6} -- \ref
{17134fg8}}
that 
HN$^{13}$C and H$^{13}$CN have a spatial distribution closer to that
of the dust emission and we
conclude that
these two isotopologues may often be very useful tracers of kinematics 
as N$_{2}$H$^{+}$ at
densities above the critical values at which CO depletes. It is possible that
CO, while depleted by roughly an order of magnitude relative to the
canonical [CO]/[H$_{2}]$ abundance ratio of 10$^{-4}$, is nevertheless
sufficiently abundant to account for minor species such as HCN and
HNC.

Finally, 
we note from Fig.~\ref{17134fg8} 
that towards \object{TMC 2}, where the dust peak
H$_{2}$ column density is larger than in the other two sources, there seems
to be a good general accord between the dust emission and the intensity of
HN$^{13}$C, H$^{13}$CN, 
and \nnh. The CO however has again a flat distribution
suggesting once more that it is depleted in the vicinity of the dust
peak. There is no reason in this source to suppose that \nnh\ traces
the high density gas around the core nucleus better than the $^{13}$C
substituted isotopologues of HCN and HNC. There are however slight
differences close to the dust peak between the different species
which are perhaps attributable to excitation and optical depth effects,
but could also be caused by the depletion which presumably all
molecules undergo if the density is sufficiently high.
We conclude
that cyanides and isocyanides are useful tracers of gas with densities
around $10^5$ cm$^{-3}$.

\begin{figure}[]
\begin{center}
\resizebox{\hsize}{!}{\includegraphics[angle=0]{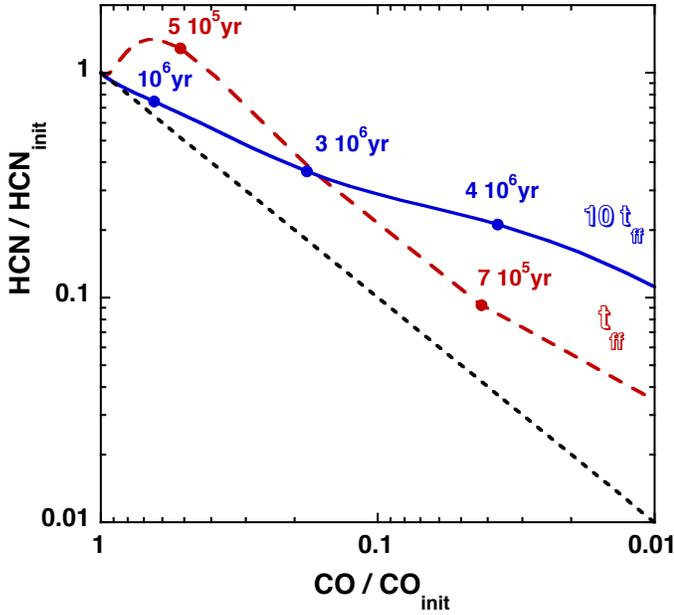}}
\caption{The HCN depletion as a function of the CO depletion during the collapse,
computed by the gravitational collapse model of Hily-Blant et al. (\cite{hw10}), see
their Fig.~8. Values are normalised with respect to the initial abundances.
Each point is labelled with the time of the collapsing gas. Two models
are shown: a free fall model labelled $t_{\rm ff}$ ({\em red dashed curve}) and a
collapsing model with a time scale multiplied by a factor ten, labelled 
$10t_{\rm ff}$ ({\em blue solid curve}). The {\em black dotted curve}
shows the positions of equal depletion for CO and HCN.}
\label{17134fg5}
\end{center}
\end{figure}

We note that models of a collapsing prestellar core indeed suggest that HCN and
HNC should remain with appreciable gas-phase abundance at an epoch when the CO
abundance has depleted to a few percent of the canonical value of $10^{-4}$
relative to H$_2$.
To illustrate the differential freeze out of HCN relative to that of CO found in the 
chemical models, Fig.~\ref{17134fg5} displays the depletion of HCN as a function of
the depletion of CO computed in the collapsing gas, starting from steady-state
abundances at a density of 10$^4$ cm$^{-3}$. The numerical gravitational collapse
model is described in Sect. 6.2 of Hily-Blant et al. (\cite{hw10}).
Two models are shown for two different collapse time scale: free-fall time
($t_{\rm ff}$)
and $10t_{\rm ff}$. It may be seen from Fig.~\ref{17134fg5} that the depletion of
HCN relative to that of CO strongly depends on the collapse time scale:
the model shows that HCN is depleted on longer timescales than
CO, in particular if the dynamical timescale is several free fall
times.
HNC, not shown in the plot, has the same behaviour of HCN.

\section{Line profiles}\label{HCN:lineprofiles}
\begin{figure*}[]
\begin{center}
\includegraphics[angle=-90,scale=.6]{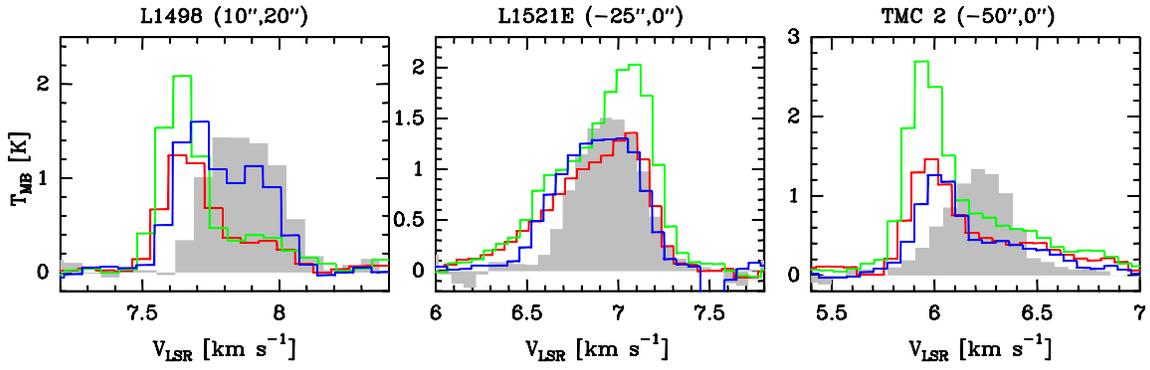}
\caption{Superposition of the three hyperfine components of
HCN(1$-$0) ($F=1\rightarrow1$, {\em red line}; $F=2\rightarrow1$, {\em green line}; 
$F=0\rightarrow1$, {\em blue
line}) with the strongest component ($F=2\rightarrow1$, {\em gray histograms}) 
of H$^{13}$CN(1$-$0) [$\times$4], see Table \ref{tab:HCN10} for component indices. 
In order to compare line shapes and intensities, components have been shifted
in frequency.}
\label{17134fg9}
\end{center}
\end{figure*}

In this Section, we discuss the behaviour of the HCN line profiles
as a function of optical depth. It is clear from Fig.~\ref{17134fg2} that towards
\object{L1498} and \object{TMC 2}, the $F=0\rightarrow1$ component
of HCN has a profile
skewed to the blue relative to $F=2\rightarrow1$ 
of H$^{13}$CN and it is tempting
to interpret this as being due to absorption in a foreground
infalling layer. 
Fig.~\ref{17134fg9} illustrates this well showing all three
HCN components compared with
the strongest $F=2\rightarrow1$ component of H$^{13}$CN. 
One sees that
there is in fact a progression going from the highly skewed optically
thick $F=2\rightarrow1$ component
of HCN to the essentially symmetric profile of H$^{13}$CN.
In the following, we attempt to quantify this trend.

Looking at Fig.~\ref{17134fg9}, related to the three hyperfine components of
HCN and the H$^{13}$CN($F=2\rightarrow1$) component, one
can qualitatively argue that
the greater the relative intensity of a line, the higher is the skewness
degree. It can be seen that the H$^{13}$CN($F=2\rightarrow1$) component
(gray 
histogram)
is fairly symmetric
and indeed, as seen earlier,
its optical depth is not high 
since it is next to optically thin limit, 
but HCN components are skewed 
towards the blue in \object{L1498} and 
\object{TMC 2}, while in \object{L1521E} they are skewed towards the red.

The superposition of the different hyperfine components of HCN
led us to the discovery of a correlation between the line profile and
its intensity. 
In particular, a red-absorbed line profile (that is skewed towards the blue) is a hint
for the presence of an outer layer which is absorbing the emission of the inner
layer while moving away from the observer, suggesting infall motions.
Conversely, a blue-absorbed line profile (that is skewed towards the red) is an
indication of the motion of the outer absorbing layer towards the observer, that is
outflow motions.

To quantify these deviations from the expected Gaussian shape in the optically thin 
limit, we compared the asymmetry degree towards the different positions mapped, using
the definition of skewness, $\delta V$, given by Mardones et al. (\cite{mm97})
\be
\delta V=\frac{V_{\rm thick}-V_{\rm thin}}{\Delta V_{\rm thin}}\,,
\ee
where $V_{\rm thick}$ and $V_{\rm thin}$ are the velocities at the peak of the optically
thick and thin components, respectively,
while $\Delta V_{\rm thin}$ is the line width of the thin
component. The normalisation of the velocity difference with $\Delta V_{\rm thin}$
reduces bias arising from lines of different width measured in different sources,
allowing a more realistic comparison of the values of $\delta V$ in our sample.
Velocities and line widths have been determined by Gaussian fits; in the event that
optically-thick line profiles were double peaked, we fitted these two components
with two Gaussians, assigning to $V_{\rm thick}$ the central
velocity of the Gaussian relative to the stronger of the two peaks.
We supposed the $F=2\rightarrow1$
component of H$^{13}$CN to be optically thin, so that
$V_{\rm thin}=V_\mathrm{H^{13}CN}$ and $\Delta V_{\rm thin}=\Delta V_{\mathrm{H^{13}CN}}$.
A value of $\delta V$ lower than zero suggests a red absorption of the line, while a 
positive value a blue absorption.

\begin{figure*}[]
\begin{center}
\includegraphics[angle=0,scale=1]{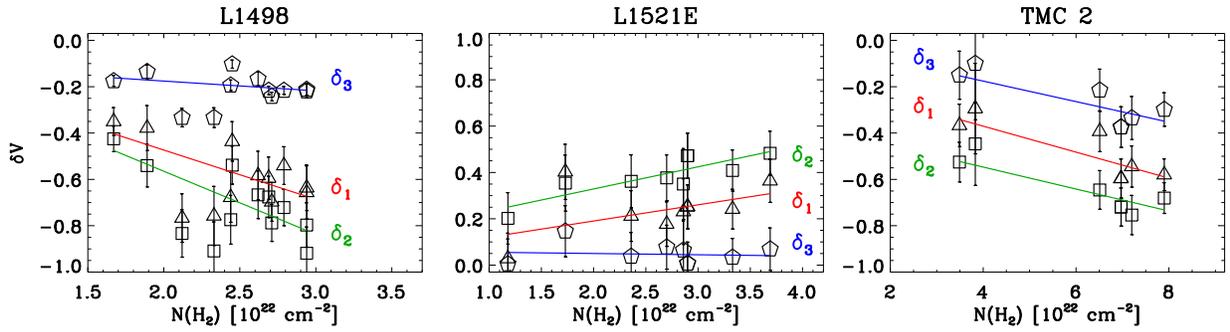}
\caption{Degree of skewness as a function of the column density of molecular hydrogen:
$\delta_{i}=(V_{i}-V_\mathrm{H^{13}CN})/\Delta V_\mathrm{H^{13}CN}$,
where $i=1,2,3$ is the $i$th component of HCN(1$-$0), and $V_\mathrm{H^{13}CN}$ and
$\Delta V_\mathrm{H^{13}CN}$ are the velocity and the line width of component 2 
($F=2\rightarrow1$, the
strongest) of the isotopologue H$^{13}$CN(1$-$0). Values of $\delta V$ from observations
of component 1 ($F=1\rightarrow1$), {\em triangles};
component 2 ($F=2\rightarrow1$), {\em squares};
component 3 ($F=0\rightarrow1$), {\em pentagons}. See also Table \ref{tab:HCN10} for component indices.}
\label{17134fg10}
\end{center}
\end{figure*}
\begin{figure*}[]
\begin{center}
\includegraphics[angle=0,scale=1]{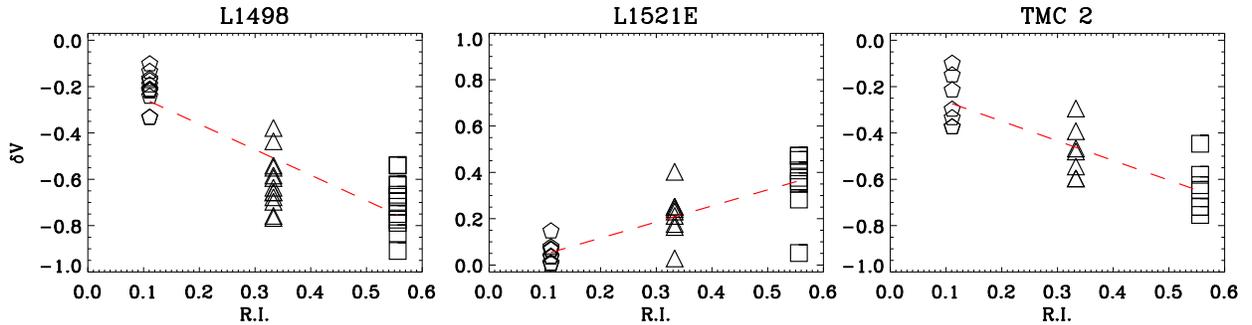}
\caption{Degree of skewness as a function of the relative intensities of the hyperfine
components of HCN(1$-$0).
Linear regressions emphasise the relationship between
$\delta V$ and line strength ({\em dashed red lines}).
Values of $\delta V$ from observations of
the $F=1\rightarrow1$ component ({\em triangles}),
the $F=2\rightarrow1$ component ({\em squares}), and
the $F=0\rightarrow1$ component ({\em pentagons}).}
\label{17134fg11}
\end{center}
\end{figure*}

Fig. \ref{17134fg10} shows the values of the skewness degree,
$\delta V$, for the three hyperfine components of HCN as a function of the 
column density of molecular hydrogen
for our source sample.
Hence we confirm the suggestion made earlier that: 
$(i)$ the absolute value of $\delta V$ is greater for
the strongest hyperfine components,
see also Fig.~\ref{17134fg11}, in fact, in all the three sources, 
$|\delta V(F=2\rightarrow1)|>%
|\delta V(F=1\rightarrow1)|>%
|\delta V(F=0\rightarrow1)|$;
$(ii)$ in \object{L1498} and \object{TMC 2}, the emission lines are red-absorbed, since $\delta V<0$, and
this is a hint for the presence of infall motions;
$(iii)$ in \object{L1521E}, the emission lines are blue-absorbed ($\delta V>0$), suggesting
expansion;
($iv$) as expected, $\delta V$ decreases from the centre to the outer part of the
cores, 
that is for decreasing values of $N$(H$_{2}$), dropping with line intensities, though sometimes not much;
($v$) $\delta V(F=0\rightarrow1)$ 
for HCN seems to be rather independent of the H$_{2}$ column
density, being almost constant
for \object{L1498} and \object{L1521E}, probably because this hyperfine component is the weakest line and 
it is closer to the optically thin limit.

\section{Excitation temperature results}\label{HCN:Texestimate}

Even though real deviations from LTE populations are present 
(see Appendix~\ref{nonltepop}), 
the LTE assumption is an approximation which is useful for many purposes and we 
calculated the excitation 
temperature,
$T_{\rm ex}$, from a simultaneous LTE fit of the hyperfine components in the observed 
species, using the measured intensity 
of optically
thick transitions.
The main-beam
temperature is given by
\be\label{TMB}
T_{\rm mb}= f_{\rm B}[J_{\nu}(T_{\rm ex})-J_{\nu}(T_{\rm bb})](1-e^{-\tau})\,,
\ee
where $J_{\nu}(T)=T_{0}/[\exp(T_{0}/T)-1]$ is the Planck-corrected brightness
temperature and $T_{\rm bb}=2.73$ K is the temperature of the cosmic background.
$T_{0}\equiv h\nu/k$, where $\nu$ is the transition frequency, and $h$ and $k$ 
represent Planck's and Boltzmann's constants, respectively. 
If both the source and the beam are Gaussian shaped, the beam filling factor, 
$f_{\rm B}$, is given by
\be
f_{\rm B}=\frac{\Omega_{\rm S}}{\Omega_{\rm B}+\Omega_{\rm S}}\,,
\ee
where $\Omega_{\rm B}=1.133\theta_{\rm B}^{2}$ and 
$\Omega_{\rm S}=1.133\theta_{\rm S}^{2}$ denote the solid angles covered by the 
beam and the source, respectively, while $\theta_{\rm B}$ and $\theta_{\rm S}$ are
the half-power beamwidth 
of the beam and the source, respectively, this latter 
evaluated by taking the line intensities
stronger than the 50\% of the peak value. We found $f_{\rm B}$ to be equal to 
unity for
most of the tracers, because $\Omega_{\rm S}$ is at least one order of magnitude 
greater than $\Omega_{\rm B}$, except for 
H$^{13}$CN and
HN$^{13}$C in \object{L1521E} ($f_{\rm B}=0.88$). 
The correct determination of the
excitation temperature for HN$^{13}$C and H$^{13}$CN is
fundamental. In fact, the gas density of the source sample 
($\sim10^5$ cm$^{-3}$,
see Tafalla et al. \cite{tm04}, Tafalla \& Santiago \cite{ts04}, and
Crapsi et al. \cite{cc05} for \object{L1498}, \object{L1521E}, and \object{TMC 2}, respectively) 
corresponds to a region where the $T_{\rm ex}$ of the two
isotopologues changes rapidly with density, namely between the radiation- and
collision-dominated density ranges. 
We can thus in principle use our $T_{\rm ex}$
determinations to constrain the density and temperature in the
region where the H$^{13}$CN and HN$^{13}$C lines are formed.  
Since these transitions are emitted in the region where CO is
depleted (see Sect.~\ref{HCN:depletion}), 
we gain information about the high density 
``core of
the core''. It is worth noting that our observations also
are a test of the collisional rates for HCN and HNC 
(Sarrasin et al. \cite{sa10}, Dumouchel et al. \cite{df10}). 
We illustrate this in the next section.

\subsection{Comparison of observed and expected excitation temperature}\label{RADEX}
\begin{figure}[!]
\begin{center}
\resizebox{\hsize}{!}{\includegraphics[angle=0]{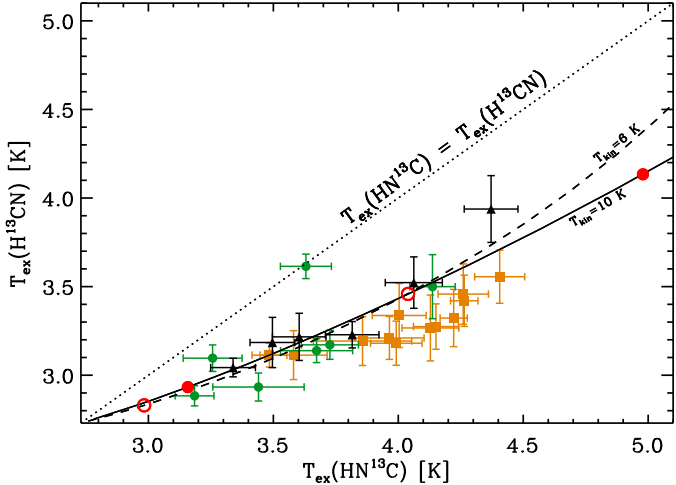}}
\caption{Excitation temperatures of H$^{13}$CN(1$-$0) versus 
HN$^{13}$C(1$-$0) evaluated
from observations of \object{L1498} ({\em yellow squares}), \object{L1521E} ({\em green circles}), and
\object{TMC 2} ({\em black triangles}). The {\em solid} and {\em dashed black lines} trace the
values for $T_{\rm ex}$ computed using RADEX for kinetic
temperatures, $T_{\rm kin}$, of 6 and 10 K; {\em red empty circles} and {\em red solid circles} show temperatures
where number density assumes values equal to 10$^4$ and 10$^5$ cm$^{-3}$ for 
$T_{\rm kin}=6$ K and 10 K, respectively.
The {\em dotted black line} 
depicts the positions where the two excitation temperatures would be equal.}
\label{17134fg12}
\end{center}
\end{figure}

In Fig.~\ref{17134fg12}, we show the comparison between the
values of the excitation temperatures of H$^{13}$CN and HN$^{13}$C evaluated from the 
simultaneous fit of the hyperfine components for the three sources examined, revealing
that $T_{\rm ex}$(HN$^{13}$C) is essentially always greater than $T_{\rm ex}$(H$^{13}$CN)
which we presume to be due to the differing
collisional rates. In the same plot, two curves trace the
values for $T_{\rm ex}$ computed using RADEX (van der Tak et al. \cite{vdtb07}) which
uses
the new collisional rates for HNC (Sarrasin et al. \cite{sa10}, Dumouchel et al. \cite{df10})
that are no longer assumed equal to HCN.
These curves assume
kinetic temperatures, $T_{\rm kin}$, equal to 6 and 10 K, and typical values of 
$2\times10^{12}$ cm$^{-2}$ and 0.2 km s$^{-1}$ for column density and line width,
respectively. 
We also assumed that the species are cospatial and hence come from regions of the
same kinetic temperature.
Notice that we rely on the fact
that these lines are not too optically thick and thus $T_{\rm ex}$ determinations are 
relatively
insensitive to column density and line width.
We notice an impressive good agreement between theory and observations, even if \object{L1498} 
shows values a little below the expected theoretical trend.

For \object{L1498}, we conclude from the RADEX results that the observed
$T_{\rm ex}$ are consistent with a density of a few times 10$^4$ cm$^{-3}$ but
this estimate is sensitive to the assumed temperature. If the kinetic
temperature is 10 K, then we can exclude densities as high as
10$^5$ cm$^{-3}$ in the region where H$^{13}$CN and HN$^{13}$C are undepleted.
On the other hand, for a temperature of 6 K, such densities are
possible. Clearly, more sophisticated models are needed to
break this degeneracy. In general, our results are consistent
with the conclusion of Tafalla et al. (\cite{tm06}) that the HCN-HNC ``hole''
is smaller than that seen in CO isotopologues. This suggests that
some CO, but much less than the canonical abundance of about 10$^{-4}$, is present in
the CO depleted region to supply carbon for HCN and HNC.

For \object{L1521E} and \object{TMC 2}, a central kinetic temperature of 6 K seems possible.
From dust emission models, they are thought to reach central densities
of around $3\times10^5$ cm$^{-3}$ (Tafalla \& Santiago~\cite{ts04}, Crapsi et al.~
\cite{cc05})
which would correspond to
excitation temperatures for H$^{13}$CN and HN$^{13}$C higher than
observed if the temperature was as high as 10 K.

\begin{figure*}[]
\begin{center}
\resizebox{\hsize}{!}{\includegraphics[angle=0]{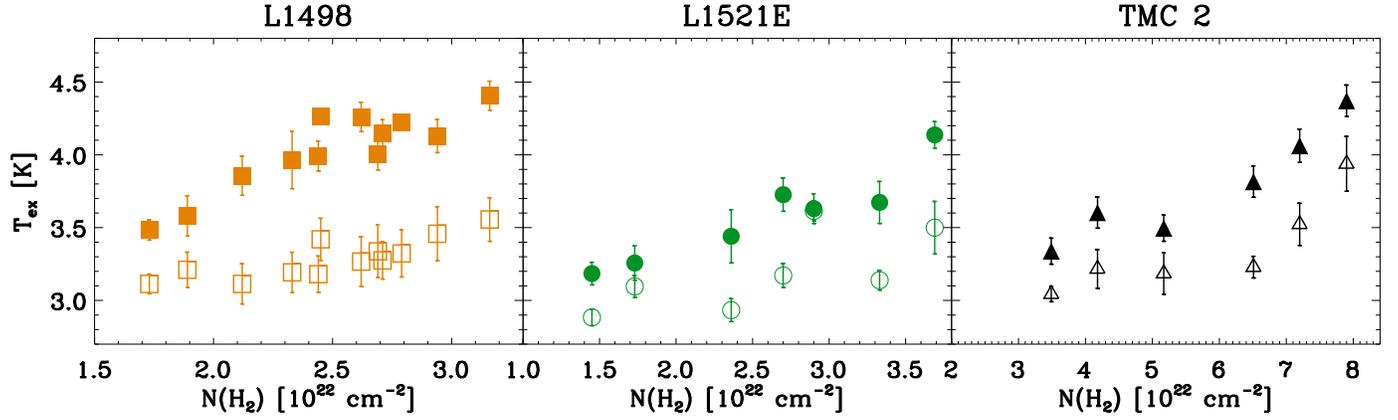}}
\caption{Excitation temperatures of H$^{13}$CN(1$-$0) and 
HN$^{13}$C(1$-$0), {\em empty} and {\em solid points}, respectively, as a function of
the molecular hydrogen column density
from observations of \object{L1498} ({\em yellow squares}), \object{L1521E} ({\em green circles}), and
\object{TMC 2} ({\em black triangles}).}
\label{17134fg13}
\end{center}
\end{figure*}

In Fig.~\ref{17134fg13}, 
we show the observed excitation temperature of
HN$^{13}$C and H$^{13}$CN 
against the H$_{2}$ column density inferred from
the dust emission. In all the three sources, one notes a clear
correlation between $T_{\rm ex}$ and $N$(H$_{2}$).  
This is an important confirmation
of the hypothesis that $T_{\rm ex}$ is a measure of collisional rates and
that the dust emission peak is also a peak in hydrogen number
density. It is also noteworthy that in \object{TMC 2}, which has the largest
central column density of our three sources, the excitation
temperatures continue to increase up to column densities of $8\times10^{22}$ cm$^{-2}$.
This suggests that HN$^{13}$C and H$^{13}$CN are still present in the gas phase
at densities close to the maximum value in \object{TMC 2} ($3\times10^5$ cm$^{-3}$
according to Crapsi et al. \cite{cc05}). On the other hand, the $T_{\rm ex}$ values at
the dust peak of \object{TMC 2} are not greatly different from that measured
at the dust peak of \object{L1498} suggesting perhaps that the central
temperature is lower in \object{TMC 2}. In any case, we conclude that
excitation temperature determinations of H$^{13}$CN and HN$^{13}$C are a
powerful technique for investigating physical parameters towards the
density peaks of cores.

\subsection{Monte Carlo treatment of radiative transfer in \object{L1498}}

\object{L1498} has been already extensively studied in molecular lines
(Tafalla et al. \cite{tm04}, \cite{tm06}, Padovani et al. \cite{pw09}) and we used 
the Monte Carlo radiative transfer code in Tafalla et al. (\cite{tm02}) to model
H$^{13}$CN and HN$^{13}$C exploiting the recent collisional rate calculations
(Sarrasin et al. \cite{sa10}, Dumouchel et al. \cite{df10}).
The core model is the same as the one used in the analysis of the molecular survey 
of \object{L1498} published in Tafalla et al. (\cite{tm06}). 
The core has a density distribution 
derived from the dust continuum map and an assumed constant gas temperature of 10 K, 
as suggested by the ammonia analysis. The radiative transfer is solved with a slightly 
modified version of the Monte Carlo model from Bernes (\cite{b79}).
The molecular parameters were taken from the
LAMDA\footnote{\tt http://www.strw.leidenuniv.nl/$\thicksim$moldata/.} database, where
the rates are computed for 
collisions with He and we assumed that the H$_{2}$ rates are larger than the He rates
by a factor of 2, the same criterion being used for HCN in Tafalla et 
al.~(\cite{tm06}).

\begin{figure}[]
\begin{center}
\resizebox{\hsize}{!}{\includegraphics[angle=0]{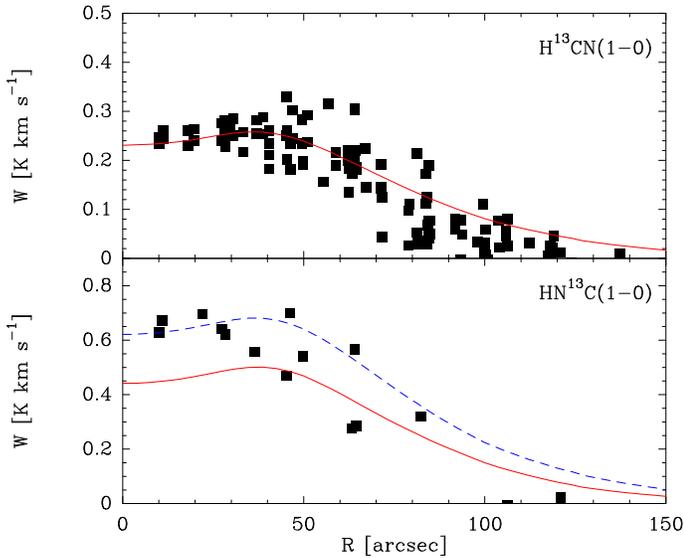}}
\caption{Radial profile of observed H$^{13}$CN(1$-$0) and HN$^{13}$C(1$-$0) 
integrated intensities 
(upper and lower panel, respectively) and model prediction for 
a core with a central hole of $8\times10^{16}$ cm and
an outer abundance 
value of $1.7\times10^{-10}$ ({\em red solid lines}) and  $2.04\times10^{-10}$
({\em blue dashed line}). HN$^{13}$C(1$-$0) 
data come from this study and H$^{13}$CN(1$-$0) data
come from this study and Tafalla et
al. (\cite{tm06}).}
\label{17134fg14}
\end{center}
\end{figure}

With regard to H$^{13}$CN,
Tafalla et al. (\cite{tm06}) made use
of the collisional rates of Monteiro \& Stutzki (\cite{ms86}) which consider
hyperfine structure complemented for higher transitions with those of Green \&
Thaddeus (\cite{gt74}) for higher $J$. In this model, the abundance law is a step
function with an outer value of $1.7\times10^{-10}$ and a central hole of
$8\times10^{16}$ cm.
We ran a model with the same abundance law, but this time using the
new LAMDA 
file for HCN, which does not account for 
hyperfine structure. 
To compensate for the lack of hyperfine structure, which spreads the photons in 
velocity, we broadened 
the line by increasing the amount of turbulence from 0.075 km s$^{-1}$ to 0.2
km s$^{-1}$. 
The upper panel of Fig.~\ref{17134fg14} shows this alternative model which nicely fits
the radial profile of observed H$^{13}$CN intensities and this 
allows us to conclude that the use of the LAMDA file for HCN without
hyperfine structure has little effect on the abundance determination, the result
being as good as that of Tafalla et al. (\cite{tm06}).

For HN$^{13}$C, we used the collisional rates for the main 
isotopologue, HNC.
As for H$^{13}$CN, the HNC hyperfine structure is not taken into account, 
so we set again the parameter relative to the turbulent velocity to 0.2 km s$^{-1}$. 
The lower panel of Fig.~\ref{17134fg14} shows the observed HN$^{13}$C 
integrated intensities
for two fits: one assumes equal abundances for H$^{13}$CN and HN$^{13}$C, the other
assumes that the HN$^{13}$C abundance is 1.2 times that of H$^{13}$CN. These
radial profiles suggest an abundance ratio between the two molecules
close to 1, with a likelihood of being closer to 1.2 towards the centre.

\begin{figure}[]
\begin{center}
\resizebox{\hsize}{!}{\includegraphics[angle=0]{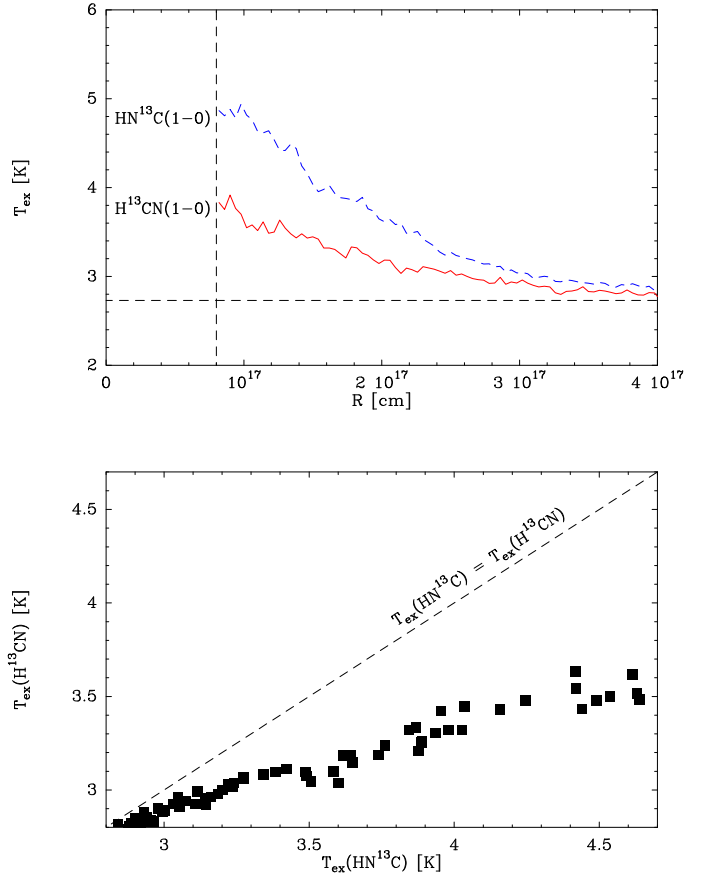}}
\caption{Upper panel: radial profile of excitation temperature for 
H$^{13}$CN(1$-$0), {\em red solid line}, and HN$^{13}$C(1$-$0), {\em blue 
dashed line}, in \object{L1498} 
as predicted by our best 
fit Monte Carlo model; the {\em vertical dashed line} represents the central
hole radius and the {\em horizontal dashed line} shows the cosmic background
temperature limit. Lower panel: comparison between HN$^{13}$C(1$-$0) and 
H$^{13}$CN(1$-$0) local values of the excitation temperature in each core shell;
the {\em dashed line} depicts the positions where the two excitation temperatures
would be equal.}
\label{17134fg15}
\end{center}
\end{figure}

The Monte Carlo code also computes the radial profile of excitation temperature, which 
is presented in the upper panel of
Fig.~\ref{17134fg15}. We notice that $T_{\rm ex}$ for both species
decreases with the radius from about 4.8 K and 3.8 K for HN$^{13}$C 
and H$^{13}$CN, respectively,
in the core interior, towards the cosmic background temperature near the outer edge of
the core. Even more interesting, we found $T_{\rm ex}$(HN$^{13}$C) 
systematically higher than
$T_{\rm ex}$(H$^{13}$CN), in nice agreement with the values computed from observations
(see lower panel of Fig.~\ref{17134fg15} compared with 
Fig.~\ref{17134fg12}). It is important to stress that 
$T_{\rm ex}$ values from
observations refer to some kind of weighted mean along the line of sight
for different positions, while $T_{\rm ex}$ values from Monte Carlo analysis are
related to each shell in the core. This means that the two numbers are 
closely connected, but do not exactly have the same meaning and a 
proper comparison would require simulating the hyperfine
analysis in the Monte Carlo spectra. 
However, the effect is global over the cloud and affects every layer,
so it cannot be ignored in any analysis. 

\section{Column density and abundance estimates}\label{HCN:Nx}
Based on our
earlier discussion, HCN hyperfine components are too optically thick
to be used for column density determinations.
This means that, while H$^{13}$CN, as well as HN$^{13}$C, originate 
in the central part
of the core, HCN is dominated by the foreground layer emission and hence it is not
possible to observe the centre of the core.
Besides, there is an unequivocal difference between the H$^{13}$CN($F=2\rightarrow1$)
and the HCN($F=0\rightarrow1$) profiles (see Fig.~\ref{17134fg9}). 
In fact, H$^{13}$CN seems to be much more
Gaussian than the weakest HCN line and it is likely to underestimate the optical
depth of HCN.

\begin{figure}[!htbp]
\begin{center}
\resizebox{\hsize}{!}{\includegraphics[angle=0]{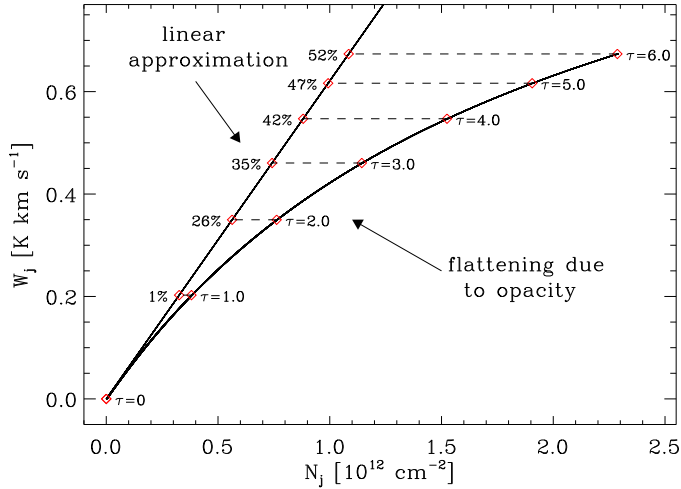}}
\caption{H$^{13}$CN(1$-$0) integrated intensity as a function of the 
column density of the $J=1$ level for 
$T_{\rm ex}=3.56$ K (the highest excitation temperature observed in \object{L1498}). 
The percentage values show the
deviation of the linear approximation with respect to the correct value taking 
account for optical depth.}
\label{17134fg16}
\end{center}
\end{figure}

However, we found that even HN$^{13}$C and H$^{13}$CN are moderately optically thick
(see Sect.~\ref{HCN:obsresults}) and in this case one can derive the column
density in the $j$th level, $N_{j}$, integrating Eq.~(\ref{TMB}) over frequency
to obtain the integrated intensity, $W_{j}$,
with the optical depth given by
\be\label{tau}
\tau=\frac{c^{3}}{8\pi\nu_{ji}^{3}}A_{ji}N_{j}(e^{h\nu_{ji}/kT_{\rm ex}}-1)\ \phi(\nu)\,,
\ee
where $A_{ji}$ is the Einstein coefficient, and $\phi(\nu)$ the profile function,
which is a sum of Gaussians (assuming a maxwellian distribution of the particle
velocities) with the appropriate weights and shifts with respect to the 
central frequency, properly accounting for the hyperfine structure. 
As shown in Fig.~\ref{17134fg16}, for small optical depths ($\tau<1$)
$N_{j}$ is in direct ratio to $W_{j}$ while, 
as $\tau$ increases, the curve flattens even if the flattening is not very sharp
because when
the main component is thick, the satellites are still thin.
Allowing for increasing optical depth is important not to underestimate
column density: 
as an instance, at the emission peak of H$^{13}$CN in \object{L1498}, where
$\tau=4.58$ (see Sect.~\ref{HCN:obsresults}), the linear approximation causes an
inaccuracy of about 45\% 
and this percentage mainly depends on the optical depth, but not on $T_{\rm ex}$. In 
fact, for decreasing excitation temperatures, the linear approximation decreases its 
slope, but also the flattening due to opacity is accentuated.
HN$^{13}$C indicates a similar deviation.

Hence, 
we determine the column density $N_{j}$ corresponding to the observed $W_{j}$ 
carrying out this procedure using the maximum and the minimum value of the
excitation temperature in order to estimate errors in $N_{j}$.
It is important to remark that in this way column densities are directly
estimated from the integrated intensity of the spectra, avoiding the use of
optical depth values evaluated from the fit of the hyperfine components which have 
often large uncertainties (see Fig.~\ref{17134fg4}).
Finally, for the total column density, $N$, it holds
\be\label{NjN}
\frac{N_{j}}{N}=\frac{g_{j}}{Q}e^{-E_{j}/kT_{\rm ex}}\,,
\ee
where 
$Q=\Sigma_{j=0}^{\infty}g_{j}\ e^{-E_{j}/kT_{\rm ex}}$ is the partition function,
$E_{j}$ and $g_{j}$ being the energy and the statistical weight of the upper 
$j$th level, respectively.

\begin{figure*}[!htbp]
\begin{center}
\resizebox{\hsize}{!}{\includegraphics[angle=0]{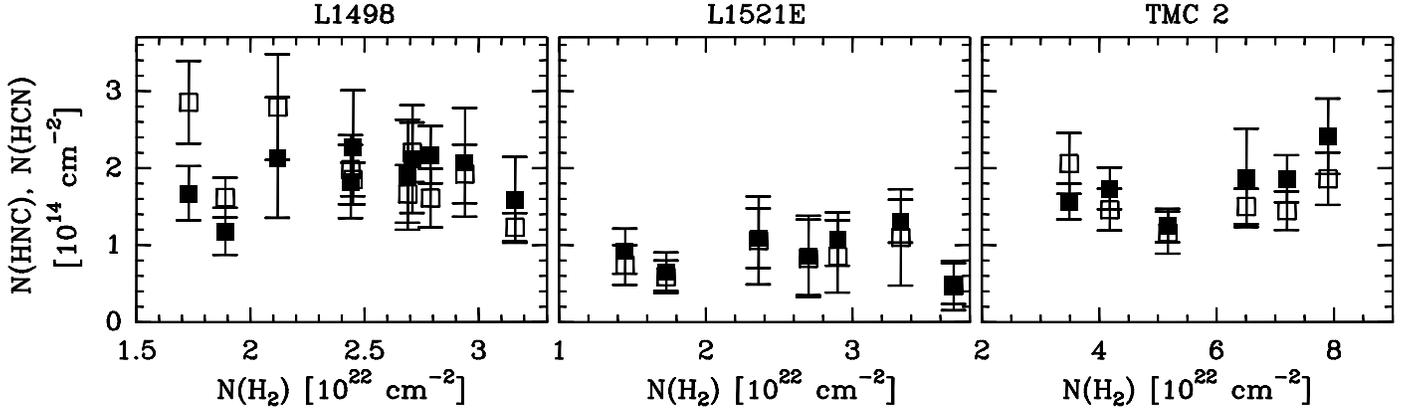}}
\caption{Column density of HNC(1$-$0), {\em solid squares}, and HCN(1$-$0), {\em empty squares} for the source sample as a function of
the H$_{2}$ column density.}
\label{17134fg17}
\end{center}
\end{figure*}

Hence, we derive the column density and the abundance of these species with
respect to molecular hydrogen for the main isotopologues HNC and HCN,
assuming the canonical isotopic [$^{12}$C]/[$^{13}$C] ratio
(Milam et al. \cite{ms05}).
Figure~\ref{17134fg17} shows HNC and HCN column densities as a function of H$_{2}$
column density which has been calculated
from the millimetre dust emission, assuming a dust
opacity per unit mass of 0.005 cm$^{2}$ g$^{-1}$ and a dust temperature of 10 K.
Notice the correlation between HNC and HCN column densities and their
optical depths in Fig.~\ref{17134fg4} for \object{L1498}: for 
$N$(H$_{2})\gtrsim2.5\times10^{22}$ cm$^{-2}$, $N$(HNC) becomes higher than
$N$(HCN) as well as $\tau(\mathrm{HNC})>\tau(\mathrm{HCN})$.
Besides, HNC and HCN abundances increase towards the peak
in \object{TMC 2} but not in the other two cores.

As an instance, in Table~\ref{coldensabund} we show the values of column density
and abundance with respect to H$_{2}$ for HNC and HCN at the dust emission peak.

\begin{table}[!h]
\caption{Column densities and abundances for HNC and HCN 
at the dust emission peak in our source sample.}
\begin{center}
\begin{tabular}{lcc}
\hline\hline
source & $N$(HNC) &  $N$(HCN)\\
& [$10^{14}$ cm$^{-2}$] & [$10^{14}$ cm$^{-2}$]\\
\hline
\object{L1498} & 1.58(0.56) & 1.24(0.18)\\
\object{L1521E} & 0.51(0.14) & 0.47(0.12)\\
\object{TMC 2} & 2.41(0.42) & 1.86(0.26)\\ 
\hline
& [HNC]/[H$_2$] & [HCN]/[H$_2$]\\
& [10$^{-9}$] & [10$^{-9}$]\\
\hline
\object{L1498} & 5.00(1.95) & 3.92(0.96)\\
\object{L1521E} & 1.36(0.42) & 1.27(0.37)\\
\object{TMC 2} & 3.05(0.57) & 2.35(0.36)\\
\hline
\end{tabular}\\[2pt]
\end{center}
\label{coldensabund}
\end{table}%

\begin{figure*}[!htbp]
\begin{center}
\resizebox{\hsize}{!}{\includegraphics[angle=0]{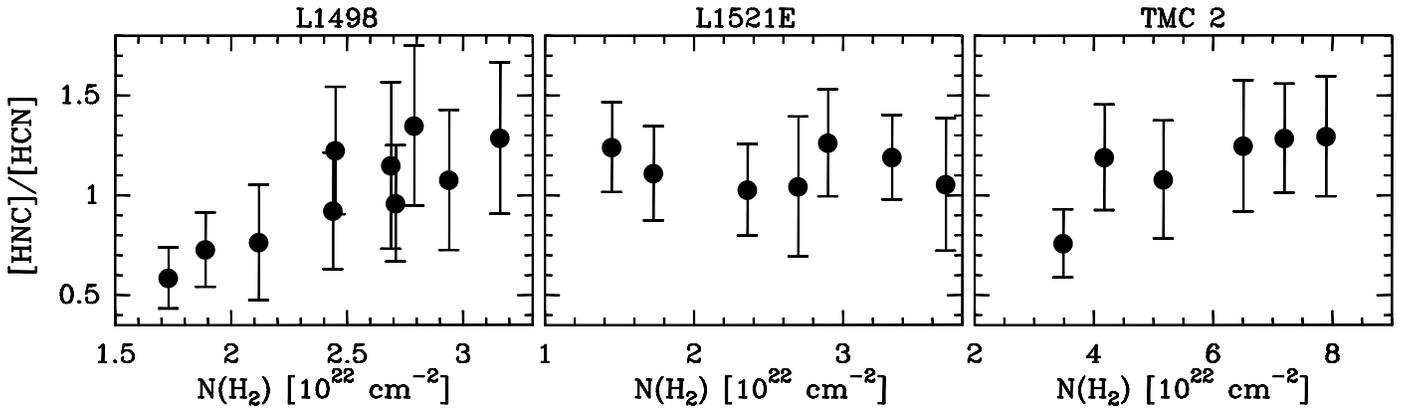}}
\caption{Abundance ratio [HNC]/[HCN] for the source sample as a function of
the H$_{2}$ column density.}
\label{17134fg18}
\end{center}
\end{figure*}

From the point of view of the chemistry, it is even more interesting to check the
ratio of abundances 
[HNC]/[HCN] as
a function of the column density of molecular hydrogen. 
Results
for the three sources of our sample are showed in Fig. \ref{17134fg18}.
We found a rather constant value for this
ratio, with weighted mean values of 
$0.94\pm0.11$ for \object{L1498}, $1.14\pm0.11$ for \object{L1521E}, and $1.05\pm0.10$ for \object{TMC 2}, 
confirming the result achieved 
by Sarrasin et al. (\cite{sa10}) and Dumouchel et al. (\cite{df10}), and 
consistent
with detailed chemical models available in the literature 
(e.g. Herbst et al. \cite{ht00}). 

This result resolves an important discrepancy
between theory and observations which has lasted almost twenty years. There
are rather few clear predictions of chemistry theory, but we can confirm that HNC and
HCN seem to have similar abundances.
However, given that what we measure is the column
density of isotopologues,
this implies that any fractionation should be the same for the two of
them.

Observations of higher-level transitions of HN$^{13}$C and H$^{13}$CN as well
as H$^{15}$NC and HC$^{15}$N would help to refine the determination of column
densities and would allow to estimate the isotopic [$^{14}$N]/[$^{15}$N] ratio, too.


\section{Conclusions}\label{HCN:conclusions}
We have studied in this article the behaviour of the $J=1\rightarrow0$
transitions of HCN(1$-$0), H$^{13}$CN(1$-$0), and HN$^{13}$C(1$-$0) as a function
of position in the three starless cores \object{L1498}, \object{L1521E}, and \object{TMC 2}.
We also observed N$_{2}$H$^{+}$(1$-$0) and C$^{18}$O(2$-$1) in \object{TMC 2}. Our main
conclusions are as follows.

\begin{itemize}
\item[1.] H$^{13}$CN(1$-$0) and HN$^{13}$C(1$-$0) are often assumed
to be optically thin when computing
column densities. Our results show that in the sources studied by us,
this is inaccurate and indeed, optical depths are sufficiently
high to make a reasonable estimate of the excitation temperatures of
these transitions which can be compared with model predictions.

\item[2.] The plot of H$^{13}$CN(1$-$0) excitation temperature 
against HN$^{13}$C(1$-$0)
excitation temperature follows the curve expected based on
the collision rates recently computed by Sarrasin et al. (2010) and
Dumouchel et al. (2010) thus confirming these results. 
This plot also stresses the importance of
calculations of potential surfaces and
collisional coefficients for isotopologues separately.
Moreover these
excitation temperatures correlate well with H$_{2}$ column density
estimated on the basis of dust emission showing that dust
emission peaks really do trace peaks in the gas density.

\item[3.] These latter results combined with our intensity-offset
plots demonstrate convincingly that, at least in \object{TMC 2}, HCN and HNC survive in the
gas phase at densities above 10$^5$ cm$^{-3}$ where CO has depleted out.
The implication of this is likely that CO survives at abundances
of a few percent of its canonical value of around $10^{-4}$ and
supplies the carbon required in lower abundance species.

\item[4.] The profiles of the three satellites of HCN(1$-$0) become increasingly
``skew'' with increasing optical depth and indeed the corresponding
H$^{13}$CN(1$-$0) profiles are reasonably symmetric. This behaviour suggests the
possibility of modelling the velocity field and abundance along the line of sight.

\item[5.] We have used a model of the density distribution in \object{L1498} to
describe the HN$^{13}$C(1$-$0) and 
H$^{13}$CN(1$-$0) results and find reasonable agreement with
a model based on the previous observations of Tafalla et al. (2006)
containing a central ``depletion hole'' of radius $8\times10^{16}$ cm.
This does not exclude models without a depletion hole, but does
confirm our conclusions on the excitation discussed above.
Indeed, rather surprisingly, our results suggest that HN$^{13}$C(1$-$0) and
H$^{13}$CN(1$-$0) trace the high density nuclei of these cores when compared
with other carbon-bearing species.

\item [6.] Our results are consistent with the models of Herbst et 
al.~(\cite{ht00})
who found that HNC and HCN should have similar abundances in prestellar
cores.

\end{itemize}

\begin{acknowledgements}
This work has benefited from research funding from the European Community's Seventh Framework Programme. We thank the anonymous referee for her/his very interesting comments that helped to improve the paper. 
\end{acknowledgements}

\begin{appendix}

\section{Non-LTE hyperfine populations}\label{nonltepop}
For a homogeneous slab with LTE between different
hyperfine levels, one has
\be
R_{ij}=\frac{1-\exp(-f_{i}\tau)}{1-\exp(-f_{j}\tau)},
\ee
where $\tau$ is the total transition optical depth 
and $f_{i}$ is the relative line strength of the $i$th component as 
in Table~\ref{tab:HCN10}.
We applied this procedure to all the transitions,
except for HN$^{13}$C(1$-$0), because of the blending of the hyperfine components.
Using the same method adopted in Padovani et al. (\cite{pw09}) for \cch, we consider a 
homogeneous slab, then a two-layer
model, and we compare different couples of ratios one versus the other to quantify
the possible departure from LTE. As for \cch, we conclude that also a two-layer
model cannot explain the observed intensities, requiring a proper non-LTE treatment.

\begin{figure}[!h]
\begin{center}
\resizebox{\hsize}{!}{\includegraphics[angle=0]{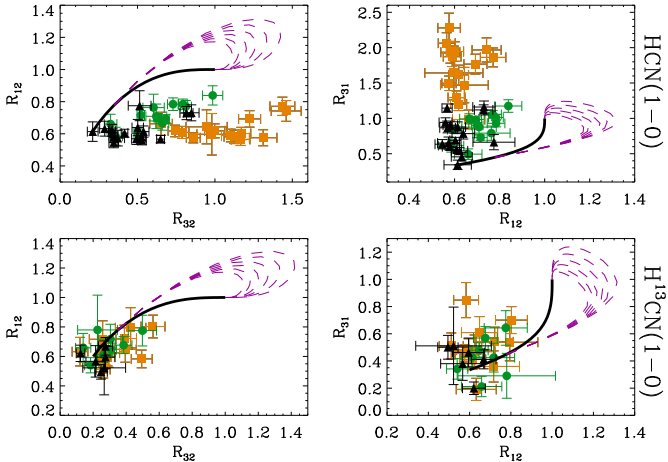}}
\caption{Ratio of the integrated intensities of couples of components of 
HCN(1$-$0) (upper panels),
and H$^{13}$CN(1$-$0) (lower panels), where $R_{ij}$ represents the ratio between 
the integrated intensities $W_{i}$ and $W_{j}$ 
(see Table \ref{tab:HCN10} for component indices).
Observational data: \object{L1498} ({\em yellow squares}), 
\object{L1521E} ({\em green circles}), 
and \object{TMC 2} ({\em black triangles}). 
One-layer model ({\em black solid curve}), 
two-layer model ({\em magenta dashed curves}).}
\label{17134fg19}
\end{center}
\end{figure}

Comparing the upper and the lower panels of Fig. \ref{17134fg19} related to HCN(1$-$0)
and
H$^{13}$CN(1$-$0), respectively, it is clear that real
departures from LTE are present and are stronger in \object{L1498} than in \object{L1521E} and
\object{TMC 2}. In \object{L1498}, $R_{32}$, the ratio between the weakest (88633 MHz) and the strongest 
(88631 MHz) component, is
larger than unity and $R_{31}>1$ in all the positions. 
This suggests a very high self (or foreground) absorption of the strongest component,
as found for the C$_{2}$H(1$-$0) emission in the same core 
(Padovani et al. \cite{pw09}).
The other two cores show
ratios which deviate from the LTE curve revealing again the presence of optical depth
effects. As an instance, in the optically thin limit, $R_{32}$ should be equal to about
0.2, but the arithmetic mean of $R_{32}$ is $\sim0.60$ 
for \object{L1521E} and $\sim0.47$ for \object{TMC 2}.

Finally, an important remark follows from the H$^{13}$CN(1$-$0) ratios
(lower panels): minor
deviations from LTE are present even in this less abundant isotopologue
which should be more optically thin. Even if optically thick lines
are the foremost responsible for the arising of non-LTE effects, the density can be
high enough to allow transition within the same rotational level 
(e.g. the transition $J,F = 1,2\rightarrow1,1$
has the same order of magnitude of rates relative to a transition between different
rotational levels, see Monteiro \& Stutzki, \cite{ms86}).

\section{Spectroscopic data and observational parameters}\label{app2}
\begin{table}[!h]
\caption{HCN(1$-$0) and H$^{13}$CN(1$-$0) frequencies of the hyperfine components
(From the JPL Molecular Spectroscopy Database: {\tt http://spec.jpl.nasa.gov}).}
\begin{center}
\begin{tabular}{cccc}
\hline\hline
Comp no. & $F^{\prime}-F$ & Frequency [MHz] & $f$\\
\hline
\multicolumn{4}{c}{HCN(1$-$0)}\\
\hline
1 & 1--1 & 88630.4160 & 0.333\\
2 & 2--1 & 88631.8470 & 0.556\\
3 & 0--1 & 88633.9360 & 0.111\\
\hline
\multicolumn{4}{c}{H$^{13}$CN(1$-$0)}\\
\hline
1 & 1--1 & 86338.7670 & 0.333\\
2 & 2--1 & 86340.1840 & 0.556\\
3 & 0--1 & 86342.2740 & 0.111\\
\hline
\end{tabular}
\end{center}
\label{tab:HCN10}
\normalsize
\end{table}%

\begin{table}[!h]
\caption{HN$^{13}$C(1$-$0) frequencies of the hyperfine components. 
For a complete list of 
the hyperfine components, see van Der Tak et al. (\cite{vdtm09}). In fact, there are four 
overlapped hyperfine components with $F_{2}^{\prime}=2$, three components with $F_{2}^{\prime}=3$, and
three components with $F=1\rightarrow1$.}
\begin{center}
\begin{tabular}{cccc}
\hline\hline
Comp no. & transition & Frequency [MHz] & $f$\\
\hline
\multicolumn{4}{c}{HN$^{13}$C(1$-$0)}\\
\hline
1 & $F_{1}'-F_{1}=0-1$ & 87090.675 & 0.065\\
2 & $\;\;F_{2}'=2$ & 87090.791 & 0.264\\
3 & $\;\;F_{2}'=3$ & 87090.834 & 0.432\\
4 & $F_{1}'-F_{1}=1-1$ & 87090.886 & 0.239\\
\hline
\end{tabular}
\end{center}
\label{tab:HN13C10}
\end{table}%

\begin{table}[!h]
\caption{Summary of observed molecules together with the observing parameters:
half power beamwidth, beam and forward efficiencies, system temperature, and 
precipitable water vapor.}
\begin{center}
\begin{tabular}{cccccc}
\hline\hline
transition & HPBW  & B$_{\rm eff}$ & F$_{\rm eff}$ & T$_{\rm sys}$ & pwv\\
& [$^{\prime\prime}$] & & & [K] & [mm]\\
\hline
HCN(1$-$0) & 28 & 0.77 & 0.95 & $\sim$130 & 1$-$2\\
H$^{13}$CN(1$-$0) & 28 & 0.77 & 0.95 & $\sim$120 & 1$-$2\\
HN$^{13}$C(1$-$0) & 28 & 0.77 & 0.95 & $\sim$140 & 1$-$2\\
N$_{2}$H$^{+}$(1$-$0) & 26 & 0.77 & 0.95 & $\sim$160 & 1$-$2\\
C$^{18}$O(2$-$1) & 11 & 0.55 & 0.91 & $\sim$320 & 1$-$2\\ 
\hline
\end{tabular}
\end{center}
\label{tab:obspam}
\end{table}%

\end{appendix}

\end{document}